\documentclass[12pt, draftclsnofoot, onecolumn]{IEEEtran}
\usepackage{graphicx}
\usepackage{epstopdf}
\usepackage{subfigure,color}
\usepackage{amssymb}
\usepackage{cite,comment}
\usepackage{amsmath}
\usepackage{bm,comment}
\usepackage{algorithm}
\usepackage{algpseudocode}
\usepackage{amsmath,epsfig}
\usepackage{epic}
\usepackage{url}

\begin{document}
\title{Safeguarding Wireless Network with UAVs: A Physical Layer Security Perspective}
\author{Qingqing Wu, Weidong Mei, and Rui Zhang
\thanks{The authors are with the Department of Electrical and Computer Engineering, National University of Singapore (NUS), Singapore 117583 (e-mails: elewuqq@nus.edu.sg, wmei@u.nus.edu, elezhang@nus.edu.sg). W. Mei is also with the NUS Graduate School for Integrative Sciences and Engineering, NUS, Singapore 119077.}}
\maketitle

\begin{abstract}
Integrating unmanned aerial vehicles (UAVs) into future wireless systems such as the fifth-generation (5G) cellular network is anticipated to bring significant benefits for both UAV and telecommunication industries. Generally speaking, UAVs can be used as new aerial platforms in the cellular network to provide communication services for terrestrial users, or become new aerial users of the cellular network served by the terrestrial base stations. Due to their high altitude, UAVs usually have dominant line-of-sight (LoS) channels with the ground nodes, which, however, pose new security challenges to future wireless networks with widely deployed UAVs. On one hand, UAV-ground communications are more prone than terrestrial communications to eavesdropping and jamming attacks by malicious nodes on the ground. On the other hand, compared to malicious ground nodes, malicious UAVs can launch more effective eavesdropping and jamming attacks to terrestrial communications. Motivated by the above, in this article, we aim to identify such new issues from a physical-layer security viewpoint and propose novel solutions to tackle them efficiently. Numerical results are provided to validate their effectiveness and promising directions for future research are also discussed.
\end{abstract}

\section{Introduction}
With flexible mobility and deployment, unmanned aerial vehicles (UAVs) or drones have found a plethora of new applications in the recent years. Typical use cases of commercial UAVs include cargo delivery, surveillance and inspection, search and rescue, aerial photography, among the others. As projected by the Federal Aviation Administration (FAA), the UAV industry will generate more than 82 billion dollars for the U.S. economy alone and create more than 100,000 new jobs in the next decade. The prosperous global market of UAVs is also envisioned to  bring new and valuable opportunities to the future wireless communication industry, such as the forthcoming fifth-generation (5G) cellular network. On one hand, to enable reliable communications for widely deployed UAVs in the future, a promising solution is to integrate UAVs into the future 5G cellular network as new aerial users served by the terrestrial base stations (BSs). In fact, recent studies by the 3rd generation partnership project (3GPP) have demonstrated the viability to support the basic communication requirements for UAVs with the existing cellular network\cite{zeng2018cellular}. On the other hand, with continuously miniaturized BSs/relays, it becomes more feasible to mount them on UAVs and make complementary aerial communication platforms in 5G cellular network to provide or enhance the communication services for the terrestrial users\cite{zeng2016wireless}.

\begin{figure}[!t]
\centering
\includegraphics[width=6in]{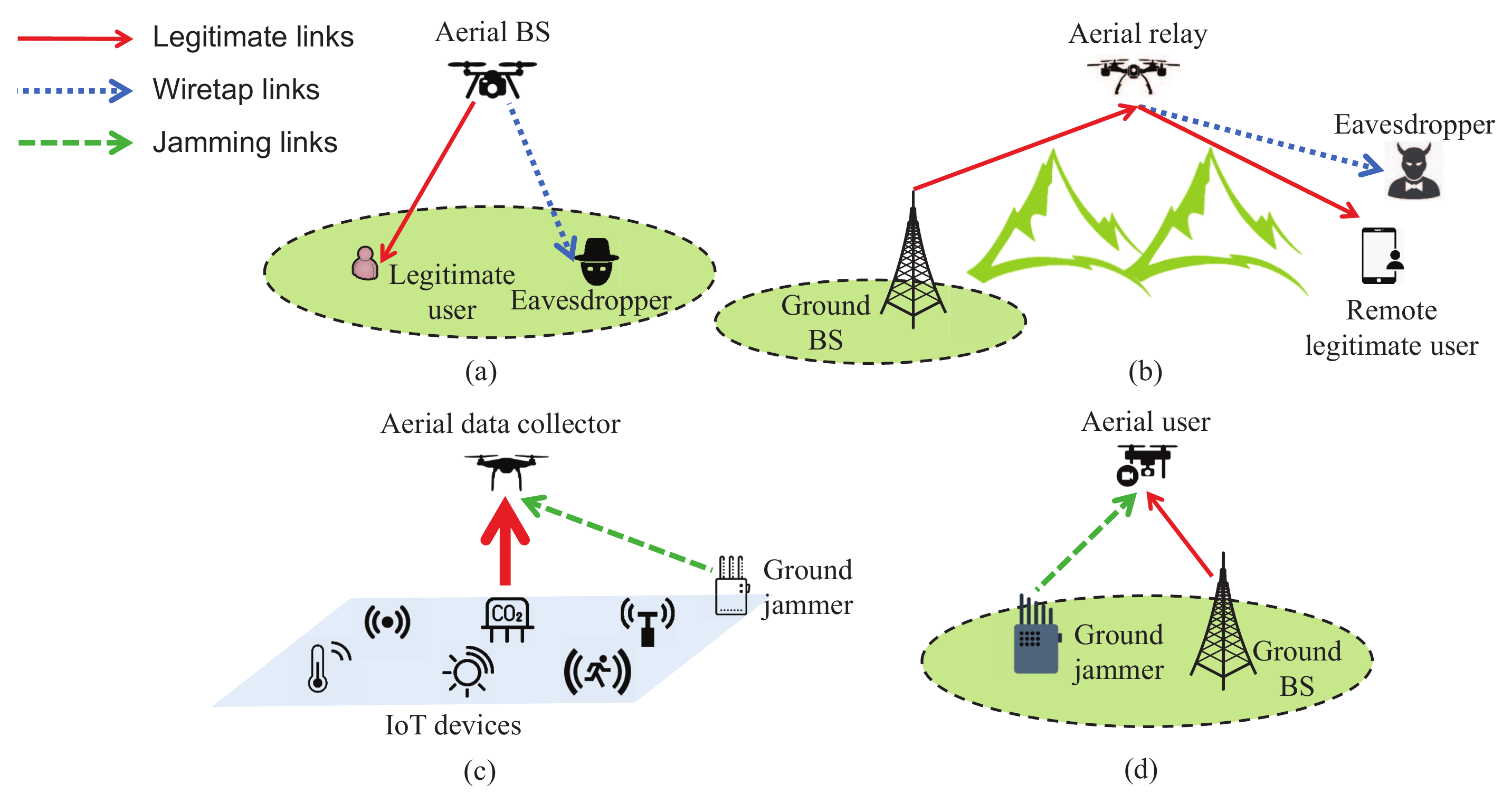}
\caption{Eavesdropping and jamming attacks to legitimate UAV communications in  5G  wireless network.}\label{UAV_Legi}
\vspace{-12pt}
\end{figure}
However, the integration of UAVs into 5G also brings new challenges that need to be addressed. Compared to terrestrial wireless channels that in general suffer from severe path-loss, shadowing, and multi-path fading, the high altitude of UAVs generally leads to more dominant line-of-sight (LoS) channels with the ground nodes. Although the strong LoS links can be exploited to improve the communication performance, they also cause severe interference to the terrestrial communications and thus effective air-ground interference management techniques are needed\cite{zeng2018cellular}. Moreover, the unique LoS-dominant UAV-ground channel poses new security challenges to the future wireless network with  various UAV applications as shown in Fig.\,\ref{UAV_Legi}. Specifically, in Fig.\,\ref{UAV_Legi}(a) and Fig.\,\ref{UAV_Legi}(b), the UAV is deployed as an aerial BS or relay that sends/forwards private messages to a legitimate ground node, respectively. However, the strong UAV-ground LoS links also enhance the reception quality of the terrestrial eavesdroppers. As a result, even a remote eavesdropper can take advantage of the LoS link to overhear the air-to-ground (A2G) communication clearly. Furthermore, in Fig.\,\ref{UAV_Legi}(c) and Fig.\,\ref{UAV_Legi}(d), the UAV acts as a mobile hub to collect private data from terrestrial Internet-of-things (IoT) devices such as sensors, or an aerial user that receives critical control signals from its serving ground BS. In such cases, a ground jammer can also exploit the strong ground-to-air (G2A) LoS links to launch more powerful jamming attacks to interfere with the UAV, thus significantly degrading its reception quality and even resulting in UAV communication failure. Therefore, compared to terrestrial communications, A2G and G2A communications are more susceptible to terrestrial eavesdropping and jamming, respectively. As such, how to effectively safeguard the legitimate UAV communications in the future wireless network is a new and challenging problem to resolve.

\begin{figure}[!t]
\centering
\includegraphics[width=6in]{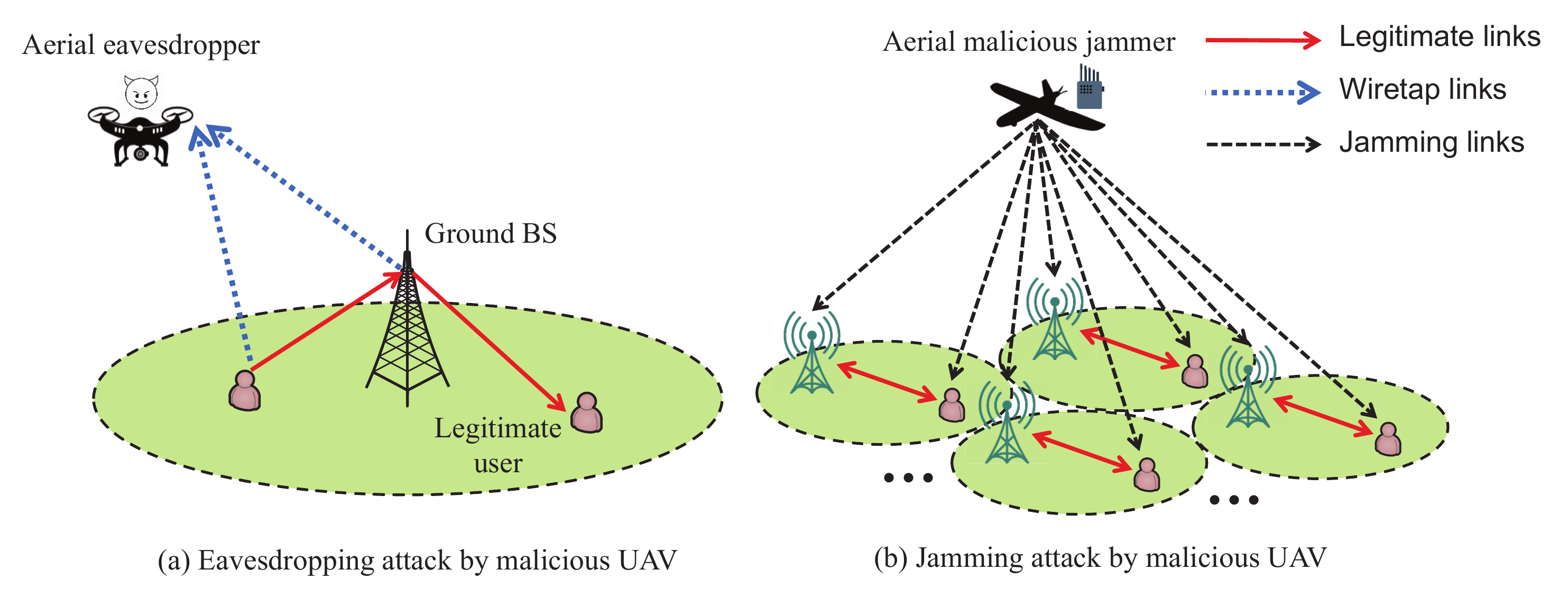}
\caption{Eavesdropping and jamming attacks to terrestrial wireless network by malicious UAV nodes.}\label{UAV_Mali}
\vspace{-12pt}
\end{figure}
On the other hand, UAVs may be a potentially new security threat to the terrestrial cellular network if they are misused by unauthorized parties for malicious purposes. As shown in Fig.\,\ref{UAV_Mali}(a), if the UAV is deployed as a malicious eavesdropper, it could more clearly overhear the terrestrial private communications between the ground BSs and their users even in a wide area as compared to the traditional terrestrial eavesdroppers, due to the more LoS-dominant G2A links. Moreover, the UAV may also be used as a malicious jammer to disrupt both the uplink and downlink terrestrial communications in the cellular network, as shown in Fig.\,\ref{UAV_Mali}(b). Its strong A2G links thus impose more severe as well as more  widely spread interference to legitimate ground communications than conventional terrestrial jammers with the same jamming signal power. In addition, the malicious UAVs can leverage their high mobility to flexibly change locations so that they can keep track of their moving targets over time and thus overhear/jam their communications more effectively. In light of the above, another challenging security problem arises, i.e., how to efficiently protect the terrestrial communications against the powerful  UAV eavesdropping/jamming attacks?

Considering the above two new UAV security issues in future wireless networks, this article aims to identify their key challenges and propose  promising solutions to tackle them, mainly from a physical-layer (PHY) design perspective. Note that PHY security has been extensively studied in the literature to combat the malicious eavesdropping and/or jamming attacks in terrestrial communication networks (see, e.g. \cite{yang2015safeguarding,jameel2018comprehensive} and the references therein). Typical anti-eavesdropping techniques include artificial noise (AN), cooperative jamming, transmit beamforming, etc. \cite{mukherjee2014principles}; while direct sequence spread spectrum (DSSS) and frequency hopping spread spectrum (FHSS) are two commonly used anti-jamming techniques at PHY\cite{mpitziopoulos2009survey}. However, such techniques consider mainly terrestrial nodes as legitimate or malicious users in the network and thus they may be ineffective to deal with the new and more challenging LoS-induced security issues with UAVs. As such, it is necessary to revisit the conventional techniques for UAV security and devise new solutions to safeguard future wireless networks more effectively.

\section{Securing Legitimate UAV Communications in Wireless Network}
As compared to terrestrial communications, legitimate A2G/G2A communications are more susceptible to ground eavesdropping/jamming due to the LoS-dominant channels. In this section, we propose promising countermeasures to improve UAV communication security and reliability.

\subsection{Protecting A2G Communication from Terrestrial Eavesdropping}
As shown in Fig.\;\ref{UAV_Legi}(a) and Fig.\;\ref{UAV_Legi}(b), for A2G communications where UAVs transmit confidential messages  to their ground nodes, the received signal strength is strong over a large area on the ground due to the LoS-dominant channels, which makes the prevention of terrestrial eavesdropping highly difficult. Nonetheless, by exploiting the high altitude of UAVs and their high mobility in the three-dimensional (3D) space, secure A2G communication may still be achievable, with the use of the following techniques.

\subsubsection{\bf UAV 3D Beamforming}
\begin{figure}[!t]
\centering
\includegraphics[width=0.65\textwidth]{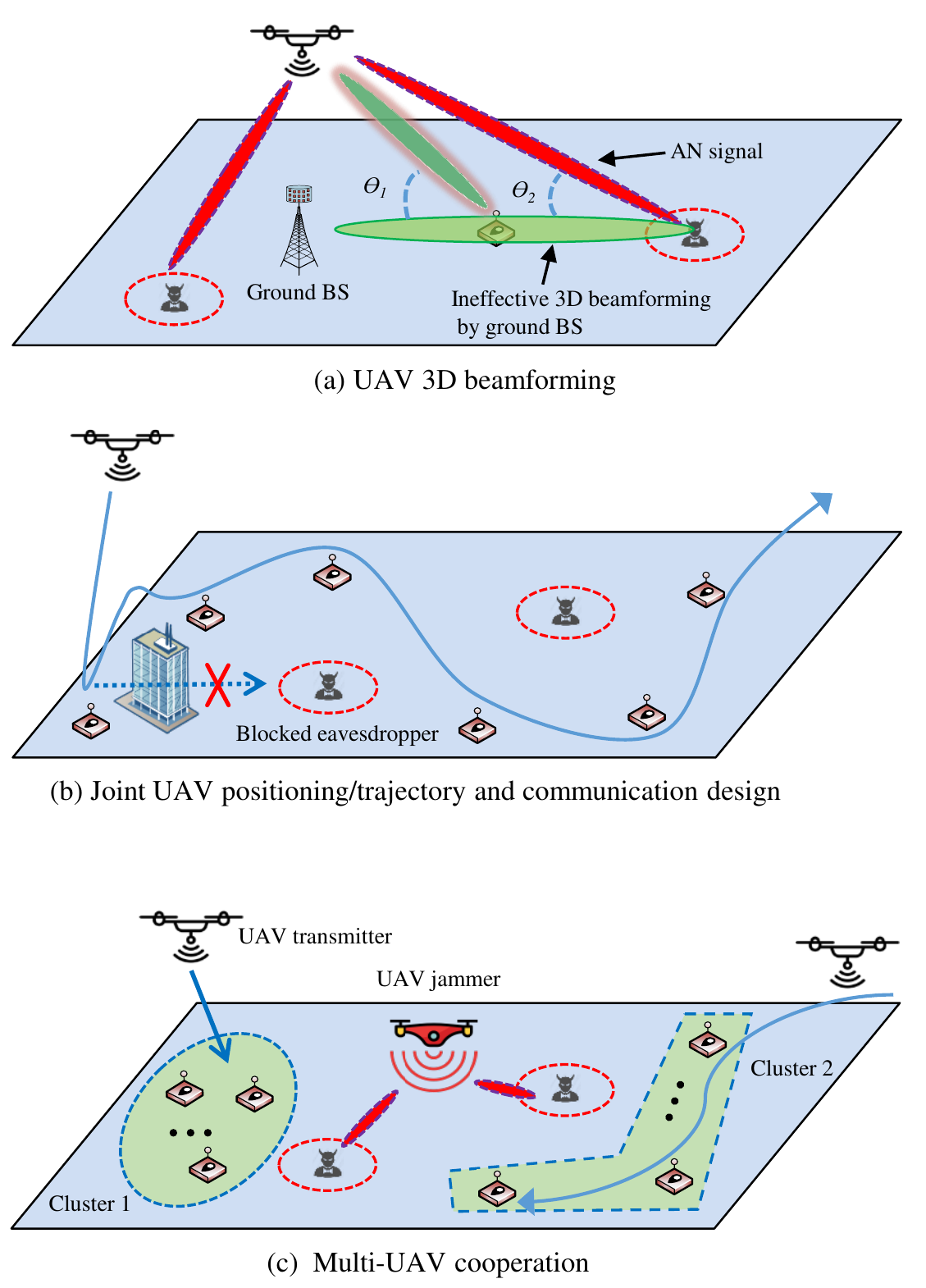}
\caption{Approaches for safeguarding legitimate A2G communications against terrestrial eavesdropping.} \label{tech1}
\end{figure}

The miniaturization of light-weight antennas and steadily increasing payload of UAVs have made it feasible to equip UAVs with multiple antennas such as planar arrays and thereby enable their 3D beamforming towards the ground, as shown in Fig.\;\ref{tech1}(a). Different from the traditional two-dimensional (2D) beamforming (e.g., with linear array) by ground BSs which controls the beam pattern only in the azimuth plane, 3D beamforming is able to adjust the beam pattern in both elevation and azimuth planes with more refined beam resolution, which can be leveraged to null the legitimate user's signal in the directions of the eavesdroppers more effectively.
Furthermore, the dominant LoS A2G channel also makes transmit beamforming more efficient as compared to that in terrestrial channels with more complicated multi-path effects due to the rich scattering environment. As shown in Fig.\;\ref{tech1}(a), even equipped with 3D beamforming capability, a ground BS may lack sufficiently separated elevation angles with the legitimate user and the eavesdropper due to its low altitude above the ground; as a result, it cannot avoid information leakage if the two nodes are located in the same direction from it, even though they have different horizontal distances with it. In contrast, a UAV transmitter at a much higher altitude can still exploit the ground nodes' different elevation angles ($\theta_1$ and $\theta_2$) with it to cancel the legitimate user's signal at the eavesdropper by e.g., zero-forcing (ZF) based precoding. Furthermore, 3D beamforming can be jointly employed with transmitting AN to jam or interfere with the ground eavesdroppers and thereby achieve better security \cite{wu2016secure}. Last but not least, the wiretap channels of the  ground eavesdroppers can be practically obtained with high accuracy at the UAV if their locations are known and thus their dominant LoS channel components can be resolved. This further enhances the robustness of 3D beamforming/jamming by the UAV.

\subsubsection{\bf Joint UAV Positioning/Trajectory and Communication Design}
As compared to terrestrial nodes, UAVs can be deployed more flexibly and move more freely in 3D space, which can be utilized to improve the communication throughput \cite{JR:wu2017joint,zeng2016throughput} or secrecy rate \cite{zhang2018securing} with the ground users.
For example, a UAV transmitter at a given horizontal position can avoid signal blockage with legitimate users due to a high-rise building by increasing its altitude, or incur blockage and thus more propagation loss with the ground eavesdropper by lowing its altitude (see Fig.\;\ref{tech1}(b)) so as to achieve better secrecy performance. On the other hand, for flying UAVs, a proper trajectory design of the UAV helps shorten its communication distances with legitimate users while enlarging the wiretap channel distances to alleviate the ground eavesdropping. Furthermore, communication scheduling and power control can be jointly designed with the UAV trajectory to achieve the optimum secrecy performance. For example, as shown in Fig.\;\ref{tech1}(b), when a UAV has to fly over some eavesdroppers due to its mission requirement, the legitimate channel may not be always stronger than the wiretap channel along its trajectory.
In this case, the UAV can increase its transmit power/rate when it flies closer to a legitimate user for communication while being further away from the eavesdroppers, and decrease transmit power/rate or even stop transmission otherwise. Finally, the UAV positioning/trajectory and communication co-design can be adaptive in real time to deal with moving eavesdroppers on the ground that intend to improve the eavesdropping performance and/or hide themselves from getting exposed.

\subsubsection{\bf Multi-UAV Cooperation}
In practice, a single UAV only has limited communication and maneuvering capability and thus may not achieve the desired secure communication performance in some challenging scenarios (e.g., with multiple collusive eavesdroppers over a large  area). This thus motivates the deployment of multiple collaborative UAVs to achieve more efficient secure communications  \cite{JR:wu2017joint,zhong2018secure}.
For example, as shown in Fig.\;\ref{tech1}(c), based on the locations of eavesdroppers, the ground users can be grouped into different  clusters with each cluster served by a single UAV. As such, it may not be necessary for the UAV to fly over the eavesdroppers as in Fig.\;\ref{tech1}(b), which thus helps reduce the information leakage. Alternatively, some UAVs may act as aerial jammers  \cite{JR:Anli18WCL,zhong2018secure}, which are deployed above a group of nearby eavesdroppers to degrade their signal reception by sending AN signals \cite{JR:Anli18WCL,zhong2018secure}. This in turn provides higher flexibility for the deployment or trajectory design of other UAV transmitters to achieve better secrecy communication performance.

\subsection{Securing G2A Communication Against Terrestrial Jamming}
In contrast to legitimate ground receivers in A2G communications, legitimate UAV receivers in G2A communications (see Fig.\;\ref{UAV_Legi}(c) and Fig.\;\ref{UAV_Legi}(d)) are exposed in the sky and thus more vulnerable to jamming attacks by ground adversary nodes that send AN signals to interfere with the UAV so as to reduce its receive signal-to-interference-plus-noise ratio (SINR) for decoding.
Although DSSS and FHSS are two widely applied anti-jamming techniques, they usually result in low spectrum efficiency and may not be sufficient to deal with the strong G2A jamming due to the LoS-dominant channel. Fortunately, the techniques proposed in the preceding subsection can also be applied to cope with the malicious jamming more effectively. For example, the 3D receive beamforming at a legitimate UAV can provide higher spatial resolution than 2D beamforming (with a linear array) and thus more efficiently cancel the interference from the ground jammers, as long as they are not so close to the legitimate ground transmitters so that their elevation angles with the UAV can be resolved. In the case that the legitimate ground transmitter is a multi-antenna BS as shown in Fig.\;\ref{UAV_Legi}(d), the 3D transmit and receive beamforming can be jointly employed at the BS and UAV to maximally improve the legitimate link SINR. Moreover, the 3D high mobility of UAV can be leveraged to optimize its position or trajectory by shortening/enlarging the distances from legitimate transmitters/malicious jammers. Finally, in the case that the UAV receiver needs to move away from the jammed area to avoid the strong interference, the device-to-device (D2D) communications among the legitimate ground nodes (e.g., the ground sensors shown in Fig.\;\ref{UAV_Legi}(c)) can be exploited to forward the data of the ground nodes in the jammed area via a separate (unjammed) channel to other nodes that are sufficiently far away from the jammer but close to the UAV position/trajectory for more reliable uploading to the UAV.

\section{Safeguarding Terrestrial Network Against Malicious UAV Attacks}
The LoS-dominant air-ground channels also make the terrestrial communications highly exposed to malicious UAVs. For example, a single UAV eavesdropper or jammer is capable of wiretapping or contaminating the transmissions within multiple cells. In addition, the flexible mobility and deployment of UAVs may be harnessed by malicious users to launch more aggressive attacks to terrestrial networks. It is therefore of paramount importance to develop advanced countermeasures for combating such attacks by malicious UAVs.

\subsection{Anti-UAV Eavesdropping of Terrestrial Communication}
As shown in Fig.\;\ref{UAV_Mali}(a), thanks to the different altitudes of a legitimate ground receiver and a UAV eavesdropper, their well separated elevation angles can be exploited by the 3D transmit beamforming at a ground BS to achieve efficient signal nulling and hence secure  communication in the downlink. However, for uplink transmissions where the legitimate ground users may not be equipped with a large  number of antennas, how to combat against malicious UAV eavesdropping becomes more challenging. To resolve this issue, we propose two promising techniques based on different cooperation mechanisms in the terrestrial network, namely multi-hop D2D relaying and cooperative remote jamming, elaborated as follows.
\begin{figure}[!t]
\centering
\includegraphics[width=0.7\textwidth]{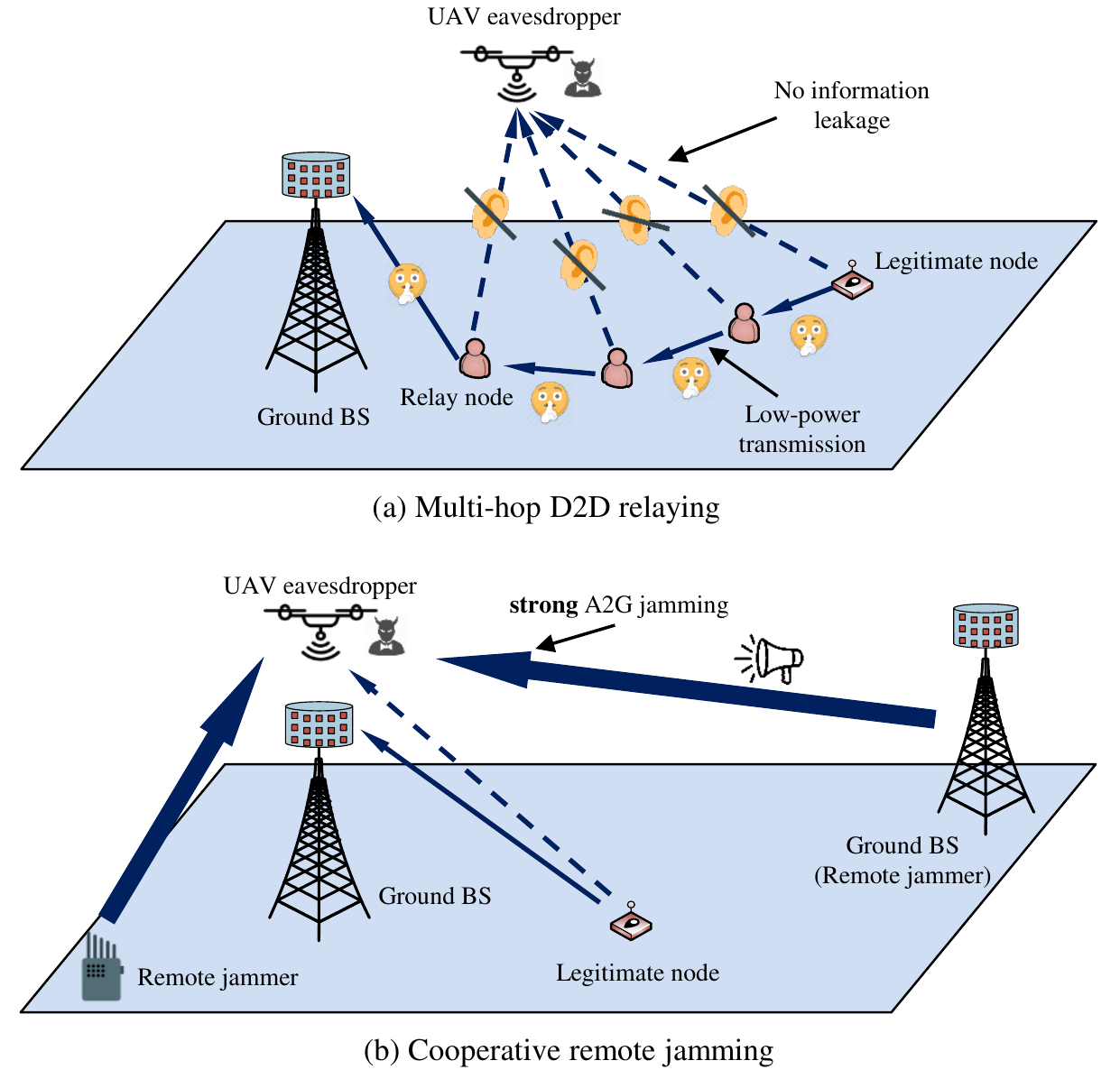}
\caption{Approaches for safeguarding terrestrial communications against UAV eavesdropping.}\label{tech2}
\end{figure}

\subsubsection{\bf Multi-hop D2D Relaying}
Given the LoS advantage of G2A channels over terrestrial fading channels, the transmitted signal that needs to be received reliably at a ground receiver in terrestrial communication is more likely to be wiretapped by the UAV under the same link distance. To overcome this difficulty, one viable solution is to adopt multi-hop D2D relaying in the terrestrial network as shown in Fig.\;\ref{tech2}(a), i.e., utilizing nearby ground nodes as legitimate relays to help forward the confidential message from a source (legitimate ground node) to its destination (ground BS). By judiciously selecting the relay nodes (e.g., those with blocked LoS paths with the UAV) and designing their cooperative transmission (e.g., transmit power control and distributed space-time coding), the spatial/multi-path diversity gain can be reaped to significantly enhance the secrecy communication rate.
However, on the other hand, an excessive number of hops may entail longer end-to-end delay as well as lower spectrum efficiency, and also increase the risk of being eavesdropped due to the more  exposure. As such, the number of hops, the transmit power in each hop, and the relay selection as well as the relay protocol need to be jointly designed to reconcile the above trade-off.

\subsubsection{\bf Cooperative Remote Jamming}
Another effective means to prevent UAV eavesdropping is the active jamming. However, its major weakness is that both the wiretap link and the legitimate link are degraded by the AN signal. To mitigate such adverse effect, a promising approach is by employing terrestrial nodes (e.g., ground BSs) at favorable locations for cooperative {\it remote} jamming (e.g., those who are close to the UAV eavesdroppers but distant from the legitimate ground receivers of interest). As shown in Fig.\;\ref{tech2}(b), a set of cooperative terrestrial BSs that are far away from the cell of the legitimate ground user  but still have strong LoS channels with the UAV eavesdropper, are selected as  jammers. As such, the signal reception at the UAV eavesdropper is dramatically degraded whereas only negligible interference is perceived at the legitimate ground receiver (serving BS) due to the more severe propagation loss and multi-path fading over terrestrial channels. In practice, depending on the traffic load  of the terrestrial network, the jamming signals of the cooperative ground BSs can be either random  message signals sent to their respective ground users or AN signals for dedicated jamming.

\subsection{Anti-UAV Jamming to Terrestrial Communication}
The jamming attack by malicious UAVs is another challenging issue for safeguarding terrestrial communications against their strong LoS jamming. Fortunately, for uplink transmissions (see Fig.\;\ref{UAV_Mali}(b)), the receive 3D beamforming can be applied at ground BSs to mitigate the UAV jamming signals efficiently. Furthermore, a neighboring idle BS that receives the jamming signal can also forward it via the high-speed backhaul link to the serving BS for cooperative jamming signal cancellation. While for downlink transmissions with single-antenna legitimate receivers (e.g., sensors), the above approaches become infeasible and anti-UAV jamming becomes more challenging.
One possible solution for this scenario  is again by exploiting the potential cooperative D2D communication of the ground nodes and their probabilistic LoS channels with the UAV jammer. Specifically, the terrestrial message is first sent to a ground node that is near to the legitimate user and also has blocked LoS link with the UAV jammer for reliable decoding. Then, the decoded message is forwarded to the legitimate user over an unjammed channel via D2D communication.

\section{Numerical Results and Discussion}
Numerical results are provided in this section to demonstrate the effectiveness of some of the proposed techniques in the previous two sections, respectively.

\subsection{UAV-assisted Jamming}
\begin{figure}[t]
\centering
\subfigure[Simulation setup.]{\includegraphics[width=0.45\textwidth]{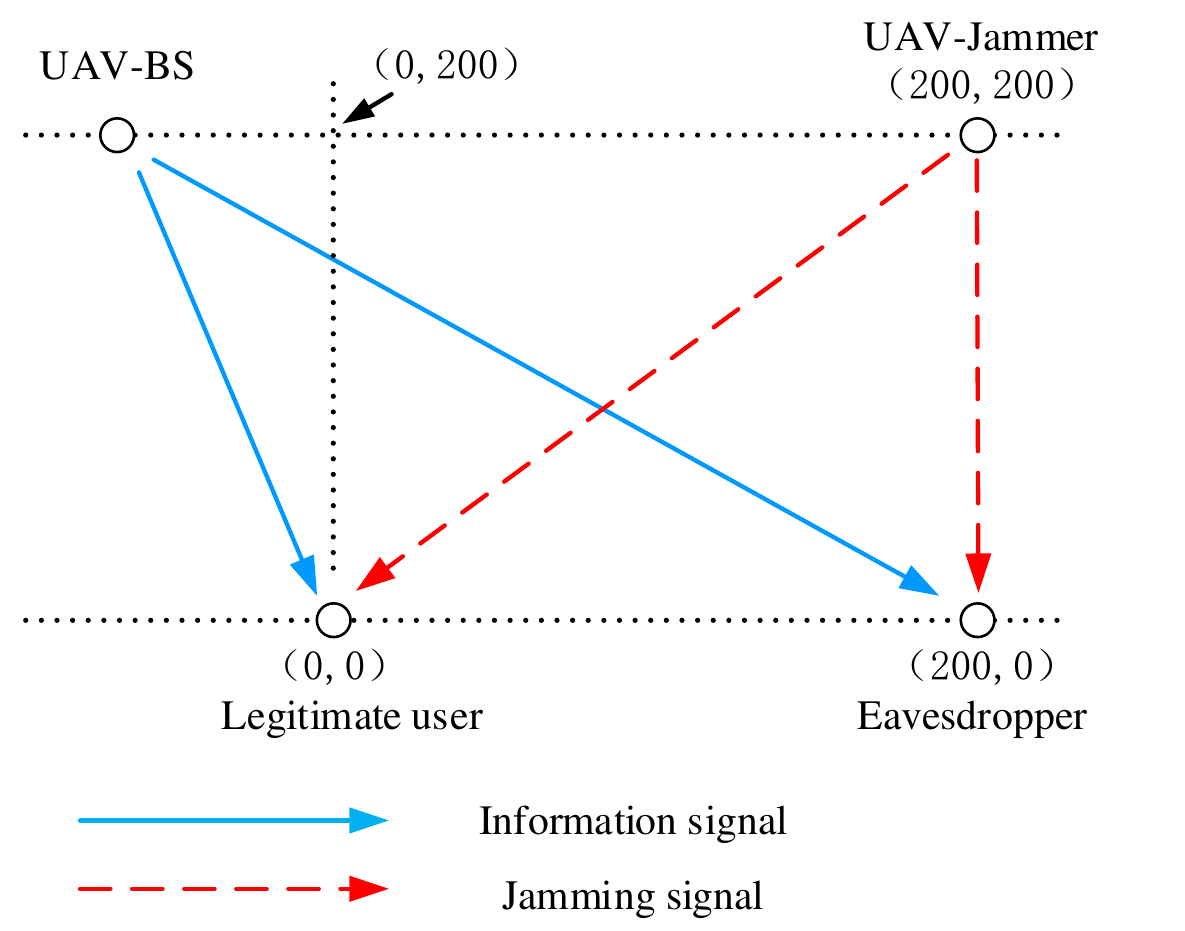}}
\subfigure[Secrecy rate versus UAV transmit power.]{\includegraphics[width=0.48\textwidth]{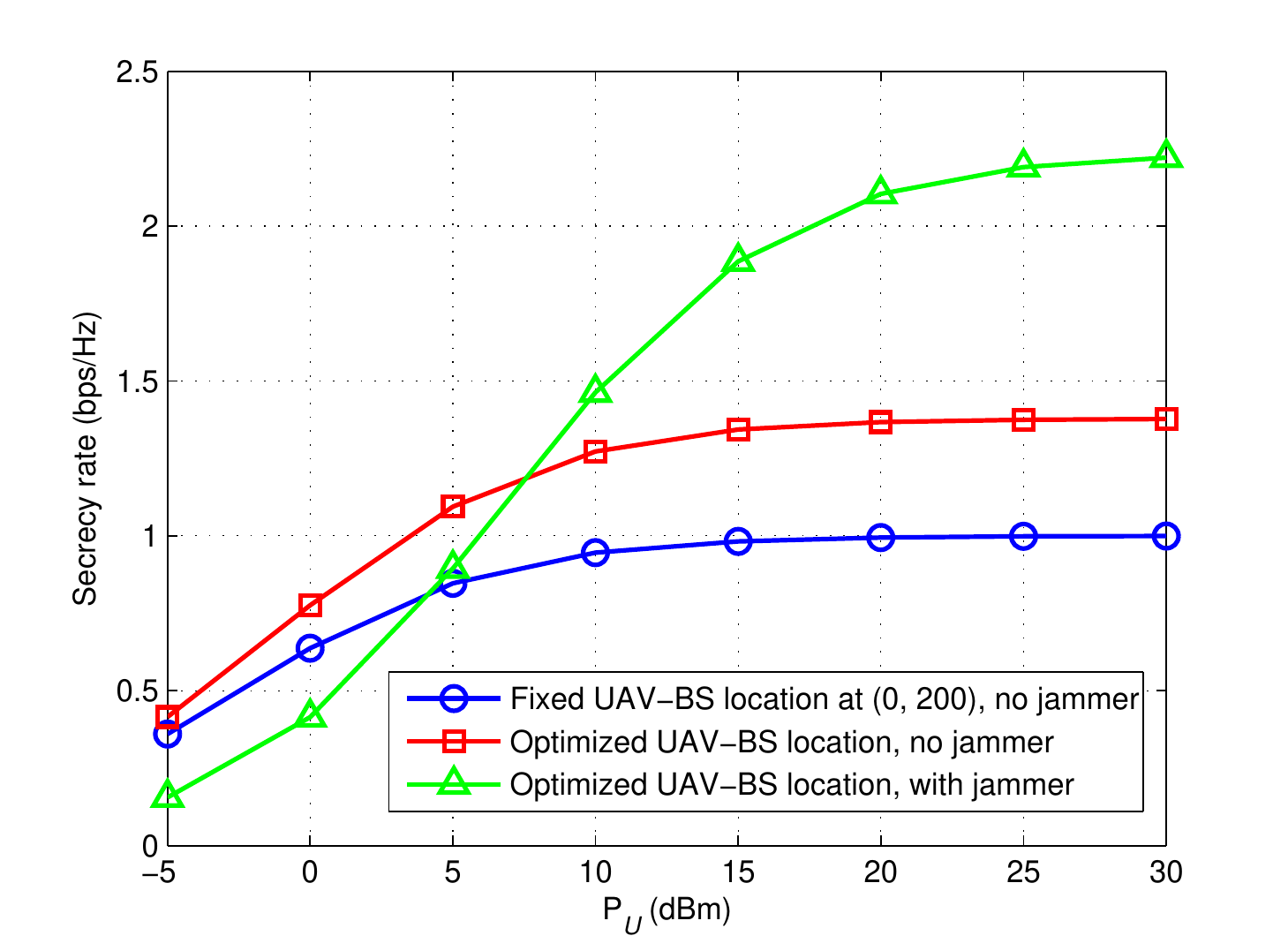}}
\caption{Secrecy rate performance with UAV-assisted jamming.}\label{multiUAV:coop} \vspace{-4mm}
\end{figure}
First, we show the performance of the proposed UAV positioning design with cooperative jamming against the terrestrial eavesdropping. As shown in Fig.\;\ref{multiUAV:coop}(a), we consider that a ground eavesdropper intends to wiretap the A2G communication from a UAV-BS to a ground user. To  improve the secrecy rate, a cooperative UAV-jammer is deployed right above the eavesdropper with the fixed jamming power given by $P_{J}=5$ dBm. The two UAVs are deployed at the altitude of $200$ meters (m) such that the A2G links approximately follow the free-space path loss model \cite{JR:wu2017joint}. The reference signal-to-noise ratio (SNR) at the distance of $1$ m is set as $80$ dB.
For comparison, we consider the following three schemes: 1) optimized UAV-BS location with jammer; 2) optimized UAV-BS location without jammer; and 3) UAV-BS location fixed  at $(0, 200)$ m without jammer, i.e., right above the legitimate user. The optimal UAV-BS locations in 1) and 2) are obtained by one dimensional search. In Fig.\;\ref{multiUAV:coop}(b), we plot the achievable secrecy rates in bits per second per Hertz (bps/Hz) by different schemes versus the UAV-BS transmit power, $P_U$. First, it is observed that compared to hovering at the fixed location, optimizing the UAV-BS locations achieves significant secrecy rate gains, even without the helping UAV-jammer. This is because the UAV-BS can enlarge the difference in the receive SNRs of the legitimate user and the eavesdropper by properly positioning itself away from both of them, i.e., hovering at the left side of $(0, 200)$. Second, one can observe that for the low UAV-BS transmit power, deploying the UAV-jammer with fixed jamming power degrades the secrecy rate; while for the high UAV-BS transmit power, the secrecy rate is substantially improved as compared to the other two schemes. This is expected as in the former case, the eavesdropping rate is quite small and hence deploying a dedicated jammer has a more detrimental effect on the legitimate receiver than the eavesdropper. While in the latter case where the eavesdropping rate becomes high, deploying the UAV-jammer in the eavesdropper's vicinity can effectively disrupt its signal reception. Meanwhile, such a cooperative UAV-jammer also enables the UAV-BS to move closer to the user for improving the secrecy rate. For example, when $P_{U}=15$ dBm, the optimized locations of the UAV-BS without and with the UAV-jammer are $(-100, 200)$ m and $(-89.5, 200)$ m, respectively. It is worth pointing out that instead of fixing the UAV-jammer's power and location in this example, we can optimize them along with the UAV-BS's location  so that the secrecy rate with the UAV-jammer is always better than that without using it \cite{zhang2018securing,zhong2018secure,JR:Anli18WCL}.

\subsection{Cooperative Remote Jamming}
\begin{figure}[hbtp]
\centering
\subfigure[Cellular network setup.]{\includegraphics[width=0.5\textwidth]{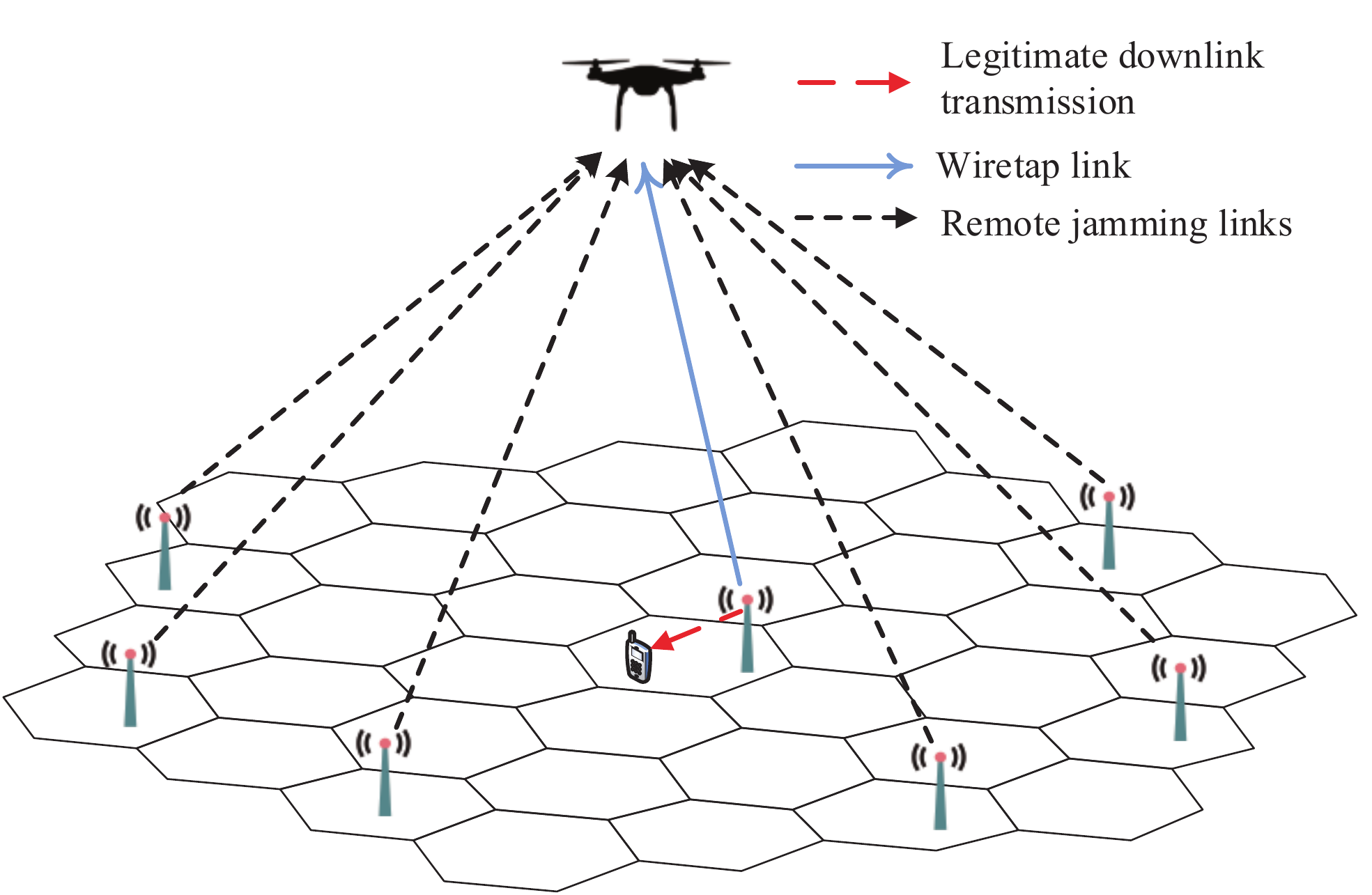}}
\subfigure[Secrecy rate versus number of cooperative BSs.]{\includegraphics[width=0.48\textwidth]{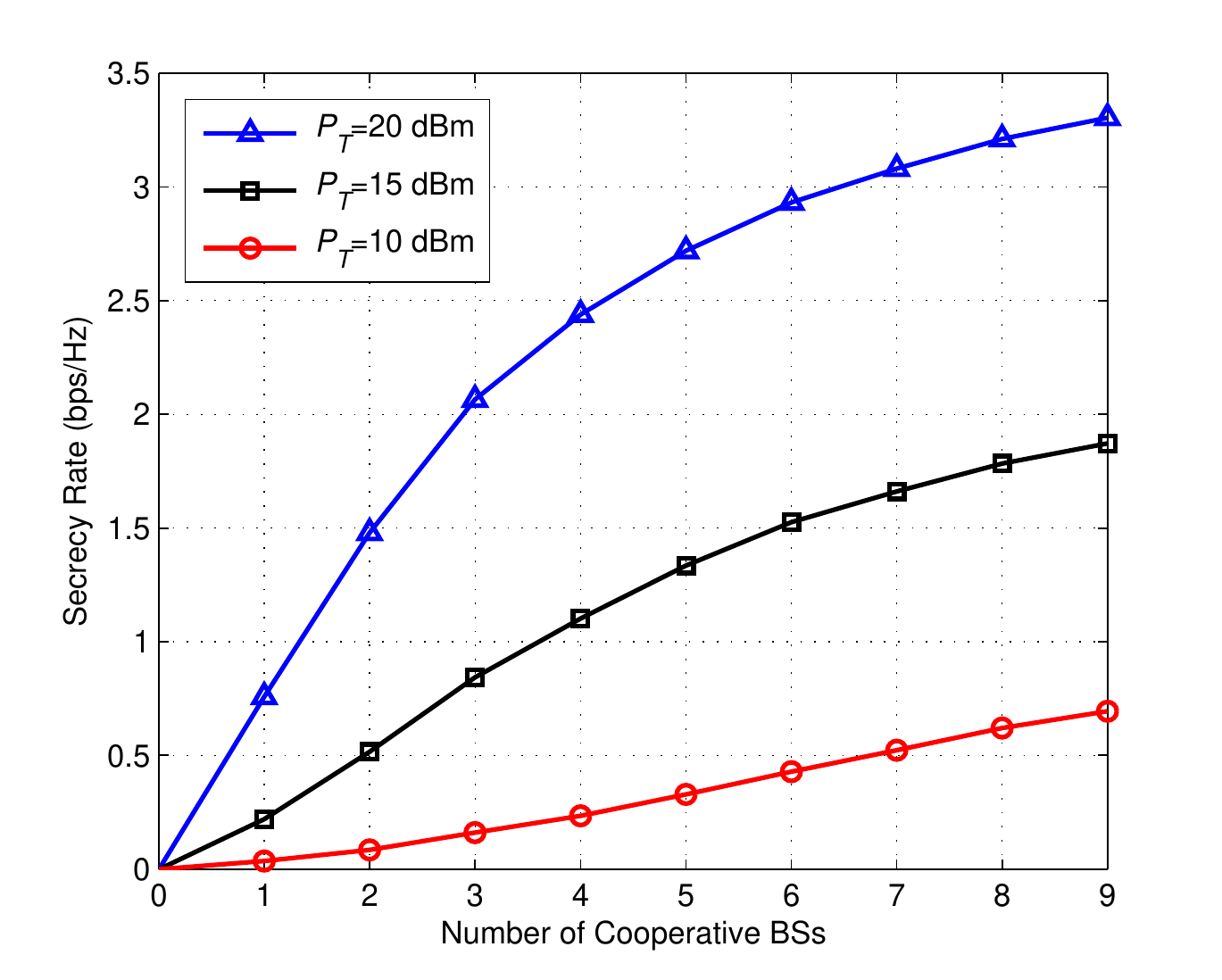}}
\caption{Secrecy rate performance with remote jamming.}\label{remote_jam}\vspace{-12pt}
\end{figure}
Next, we evaluate the performance of the proposed cooperative remote jamming technique for protecting the terrestrial communication from the UAV eavesdropping. As shown in Fig.\,\ref{remote_jam}(a), we consider a cellular network where a UAV eavesdropper aims to overhear the downlink communication in a specific cell. To combat against the UAV's strong LoS eavesdropping link, multiple remote BSs send independent jamming signals to the UAV eavesdropper. In Fig.\,\ref{remote_jam}(b), we plot the secrecy rate of the legitimate communication versus the number of cooperative BSs with  different BS transmit power, denoted as $P_T$. The cell radius is assumed to be 800 m. The UAV altitude is set to be 200 m, and the horizontal distance between the UAV and the legitimate BS transmitter is set to be 1000 m. The location of the legitimate receiver is randomly generated in the cell of the serving BS. The ground user and the UAV are assumed to be equipped with an isotropic antenna, while each BS employs a fixed antenna pattern with 10-degree downtilt angle \cite{zeng2018cellular}. From Fig.\,\ref{remote_jam}(b), it is observed that the secrecy rate is approximately  zero when there are no cooperative jamming BSs, regardless of the BS's transmit power. This is because the G2A eavesdropping channel with dominant LoS is much better than the terrestrial fading channel of the legitimate link, thus leading to virtually zero secrecy rate. However, the secrecy rate is observed to increase rapidly as the number of cooperative BSs increases. This is expected since the UAV eavesdropper suffers from more interference from the increasing number of jamming BSs, which only cause negligible interference to the legitimate receiver due to the more severe pathloss and multi-path fading over the terrestrial channels. Furthermore, it is observed that increasing $P_T$ further improves the secrecy rate, despite that the achievable rates of both the legitimate link and the UAV eavesdropping link increase with $P_T$. The reason lies in that the former increases much faster than the latter, which is severely limited by the increasing G2A interference with higher $P_T$.

\section{Conclusions and Future Work}
In this article, we focus on addressing two new and challenging security issues arising from the LoS-dominant UAV-ground channels in future wireless networks, from a PHY design perspective. We propose promising approaches and techniques to achieve secure UAV-ground communications in the presence of terrestrial eavesdroppers/jammers, as well as secure terrestrial communications against the malicious eavesdropping/jamming attacks by the UAVs. In future wireless networks, legitimate aerial and terrestrial communications both need to be protected against sophisticated attacks that may involve collusive eavesdroppers and jammers on the ground as well as over the air. As such, the proposed approaches in this article need to be further investigated to address more challenging scenarios in practice. Besides the PHY security issues discussed in this article, there are other related problems that need to be addressed, as listed below to motivate future work:
\begin{itemize}
  \item {\it UAV-assisted terrestrial adversary detection:} With the rapid advances of high-definition optical cameras, it would be an appealing solution to deploy them on UAVs for detecting, identifying, and tracking  malicious nodes on the ground. In addition, thanks to the high altitude and favorable LoS-dominant link, the UAV generally has a stronger spectrum sensing ability than conventional terrestrial nodes. Thus, the UAV can also help achieve more accurate terrestrial jamming detection.
  \item {\it Malicious UAV detection:} When there are malicious UAVs, it is imperative to detect their presence and also track their moving locations. For active UAVs such as UAV jammers, they can be detected/localized by using conventional signal sensing and ranging techniques. Whereas for passive UAVs such as UAV eavesdroppers, more sophisticated detection methods are generally required, such as radar and computer vision based methods\cite{guvencc2017detection}.
  \item {\it UAV spoofing:} In addition to the eavesdropping and jamming attacks, legitimate UAV communications may suffer from a more advanced spoofing attack such as the global position system (GPS) spoofing\cite{kerns2014unmanned}, where the ground adversary aims to deceive the UAV's navigation system by sending counterfeit GPS signals.
  \item {\it UAV-aided wireless surveillance and intervention:} By exploiting its high mobility and LoS channel with the ground, UAV can  act as a legitimate eavesdropper or jammer for realizing the surveillance of suspicious communications or intervention of malicious communications on the ground, respectively\cite{xu2017surveillance}. In contrast to the techniques proposed in this article  to defend against UAV eavesdropping and jamming, the objective is reversed in this case  to design efficient eavesdropping and jamming schemes for legitimate UAVs.
\end{itemize}

\bibliographystyle{IEEEtran}
\bibliography{IEEEabrv,mybib2}
\end{document}